\documentclass[12pt,a4paper]{article} 
\usepackage[latin1]{inputenc}
\usepackage[T1]{fontenc} 
\usepackage[english]{babel}
\usepackage[dvips]{graphics} 
\usepackage[dvips]{color} 
\usepackage{relsize}
\usepackage{graphicx,amsmath,amssymb,amsfonts} 
\title{Electromagnetic Lifshitz formula for small-width mirrors from functional determinants} 
\author{C.~D.~Fosco
and
M.~L.~Remaggi\\
{\normalsize\it Instituto Balseiro}\\ 
{\normalsize\it Universidad Nacional de Cuyo}\\ 
{\normalsize\it Centro At\'omico Bariloche}\\ 
{\normalsize\it Comisi\'on Nacional de Energ\'\i a At\'omica}\\ 
{\normalsize\it R8402AGP Bariloche, Argentina}} 
\begin{document} 
\date{} 
\maketitle 
%====================================================================
\begin{abstract} 
	We extend a recently proposed Quantum Field Theory (QFT) approach
	to the Lifshitz formula, originally implemented for a real scalar
	field, to the case of a fluctuating vacuum electromagnetic (EM)
	field, coupled to two flat, parallel mirrors.  The general result
	is presented in terms of the invariants of the vacuum polarization
	tensors due to the media on each mirror.  We consider mirrors that
	have small widths, with the zero-width limit as a particular case.
	We apply the latter to models involving graphene sheets, obtaining
	results which are consistent with previous ones.
\end{abstract}
%%%%%%%%%%%%%%%%%%%%%%%%%%%%%%%%%%%%%%%%%%%%%%%%%%%%%%%%%%%%%%%%%%%%%%%%%%
%%%%%%%%%%%%%%%%%%%%%%%%%%%%%%%%%%%%%%%%%%%%%%%%%%%%%%%%%%%%%%%%%%%%%%%%%%
%%%%%%%%%%%%%%%%%%%%%%%%%%%%% Introduction %%%%%%%%%%%%%%%%%%%%%%%%%%%%%%%
%%%%%%%%%%%%%%%%%%%%%%%%%%%%%%%%%%%%%%%%%%%%%%%%%%%%%%%%%%%%%%%%%%%%%%%%%%
%%%%%%%%%%%%%%%%%%%%%%%%%%%%%%%%%%%%%%%%%%%%%%%%%%%%%%%%%%%%%%%%%%%%%%%%%%
\section{Introduction}\label{sec:intro} 
Lifshitz' formula~\cite{Lifshitz}, provides a quite
useful tool for the evaluation of the Casimir force~\cite{Casimir} between
bodies with parallel planar interfaces, and rather arbitrary
frequency-dependent dielectric functions.  In its original
version, two disjoint media-filled half-spaces with plane, parallel
boundaries were considered; the calculation was performed at finite
temperature, and the final result for the interaction force was presented
in terms of the dielectric functions that described, macroscopically, the
electromagnetic properties of each media.  

The successive refinements achieved in precision experiments measuring the
Casimir force have provided a continuous stimulus  to generalize the scope
of the Lifshitz formula, in order to encompass either new or more realistic
situations~\cite{Reviews}. One of those generalizations has been to
consider models where the fluctuating vacuum field, rather than being
subject to ideal, `sharp and strong' boundary conditions, is instead in the
presence of background potentials, localized on the
mirrors~\cite{MIT,nos1}.  These potentials are meant to implement smooth versions of the
perfect boundary conditions. A possible way to justify them is by resorting
to the microscopic point of view. Indeed, by taking into account the
interaction of the internal degrees of freedom on the mirrors with the
fluctuating field~\cite{nos1,nos2}, one may derive an approximate effective
action for the vacuum field, containing potentials with support at the
positions of the material slabs. Even assuming them, as we shall do
throughout this paper, to have time independence and translation invariant
properties along the two `parallel' directions, \mbox{${\mathbf
x}_\parallel \equiv (x_1,x_2)$}, the potentials are, in general, nonlocal
functions of time ($x_0$)  as well as of ${\mathbf x}_\parallel$ and $x_3$.
The non locality in \mbox{$x_\parallel \equiv (x_0,x_1,x_2)$} can be dealt
with by a Fourier transformation in $x_\parallel$, since this yields a
potential which is local in frequency as well as in the  parallel
components of the momentum.  The resulting Fourier transformed potential
will still carry a dependence on the normal coordinate $x_3$, the
direction along which the effect of the potential on the fluctuating field
is strongest. The potential must be, then, necessarily non invariant under
translations in $x_3$. We shall nevertheless assume that its dependence on
$x_3$ is local~\footnote{Non localities along the normal coordinate can be
incorporated, for example, in an approach like the one
of~\cite{Fosco:2009cw}.}.

In~\cite{CcapaTtira:2011ga} a QFT approach was used to derive Lifshitz formula for a
fluctuating real scalar field coupled to two material slabs, in a situation
like the previously described one regarding both the geometry involved and
the simplifying assumptions made.  It is the aim of this article to  adapt
the approach of that reference to the case of a fluctuating Abelian gauge
field. The derivation in~\cite{CcapaTtira:2011ga} relied upon the application of
the Gelfand-Yaglom (G-Y) formula for functional determinants~\cite{GY} (for
a modern review, see~\cite{reviewGY}), objects which arise quite naturally
within the path integral formulation, for example, when incorporating
corrections due to fluctuations, in the presence a nontrivial background.

Although we shall mostly deal with zero temperature calculations, it is
convenient, for the sake of generality,  to formulate the problem in terms
of the Casimir free energy per unit area, $\Gamma_C(\beta)$. This may, in
turn, be obtained from the partition function ${\mathcal Z}(\beta)$:
\begin{equation}\label{eq:defe}
	\Gamma_C(\beta) \;= - \, \frac{1}{\beta} \, \lim_{L\to \infty} 
\Big[ \frac{1}{L^2} \, \log \frac{{\mathcal Z}(\beta)}{{\mathcal Z}_0(\beta)}\Big] \;,
\end{equation}
where $L$ is a length that characterizes the size of the plates.
${\mathcal Z}(\beta)$ can be written as an Euclidean functional integral:
\begin{equation}\label{eq:defzetabeta}
	{\mathcal Z}(\beta) \;=\; \int {\mathcal D}A \, e^{-{\mathcal
	S}(A)} \;,
\end{equation}
where ${\mathcal S}(A)$ is the Euclidean action for the gauge field, including
its coupling to the mirrors. The integral over the time-like Euclidean
coordinate $x_0$ is understood to be taken over a finite interval of length
$\beta$, with periodic boundary conditions for the field. The spatial
coordinates are assumed to be confined to a box of side length $L$, with
Dirichlet boundary conditions~\footnote{The final result, for $L \to \infty$,
shall be insensitive to the choice of boundary conditions on that spatial
box.}.  

Since we shall be interested in the Casimir force, we discard factors
independent of $l$, the distance between the mirrors. That is represented
in (\ref{eq:defe}) by the division by ${\mathcal Z}_0$, which denotes the
partition function when the  mirrors are infinitely far apart. 

Relevant physical observables shall be the vacuum energy per unit area 
\mbox{${\mathcal E}_{vac} =\lim_{\beta\to\infty} \Gamma_C(\beta)$}, as well
as the Casimir force per unit area, ${\mathcal F}_C(\beta)$:
\begin{equation}\label{eq:caf1}
	{\mathcal F}_C(\beta)\;=\; - \frac{\partial{\Gamma_C(\beta)}}{\partial l} \;,
\end{equation} 
and its zero-temperature limit \mbox{${\mathcal F}_C \equiv \lim_{\beta\to\infty}
{\mathcal F}_C(\beta)$}.

In this article, we derive expressions for $\Gamma_C(\beta)$ as a function
of the invariants that define the vacuum polarization tensor for the media
on the mirrors, as well as of the `shape' of the mirrors, understanding by
that the specific form of the $x_3$ dependence of those tensors.  We do
that for (finite) small-width mirrors and for zero-width mirrors, as an
important special case of the former.  In both cases we
consider, we take advantage of the fact that the problem is essentially
one-dimensional, and that it can be reduced to a collection of scalar
problems. For them, we apply G-Y theorem for its exact evaluation. 

This paper is organized as follows: in section~\ref{sec:themodel} we
introduce the class of model that we shall consider, writing the partition
function in terms of the physical objects that define the system: the
positions and shapes of the mirrors and their vacuum polarization tensors.
Then in section~\ref{sec:reduction}, we transform the system into two
one-dimensional scalar problems.

In~\ref{sec:Lifshitz}, we start from the partition function and show that
it can be so transformed as to be evaluated using the results
of~\cite{CcapaTtira:2011ga}. We then present the corresponding Lifshitz
formula.  

The Casimir effect for systems involving graphene sheets has been recently
studied in a series of interesting 
papers~(\cite{Bordag:2006zc},\cite{Bordag:2009fz},\cite{Fialkovsky:2011pu}),
including thermal effects. In section~\ref{sec:zerowidth} we apply the general
formula to that kind of system  as a consistency check, deriving an explicit expression for
cases involving graphene mirrors as a function of the parameters defining
the vacuum polarization tensor.
In section~\ref{sec:conclusions} we present our conclusions.
%%%%%%%%%%%%%%%%%%%%%%%%%%%%%%%%%%%%%%%%%%%%%%%%%%%%%%%%%%%%%%%%%%%%%%%%%%%%%%%
%%%%%%%%%%%%%%%%%%%%%%%%%%%%%%%%%%%%%%%%%%%%%%%%%%%%%%%%%%%%%%%%%%%%%%%%%%%%%%%
%%%%%%%%%%%%%%%%%%%%%%%%%%%%%%%  The model %%%%%%%%%%%%%%%%%%%%%%%%%%%%%%%%%%%%
%%%%%%%%%%%%%%%%%%%%%%%%%%%%%%%%%%%%%%%%%%%%%%%%%%%%%%%%%%%%%%%%%%%%%%%%%%%%%%%
%%%%%%%%%%%%%%%%%%%%%%%%%%%%%%%%%%%%%%%%%%%%%%%%%%%%%%%%%%%%%%%%%%%%%%%%%%%%%%%
\section{The model}\label{sec:themodel}
Throughout this article, we consider models where the EM field is coupled
to two imperfect mirrors modeled by `potentials' which are local in $x_3$ and
translation invariant in $x_\parallel$. Note, however, that those potentials, since
they couple to the gauge field, will also have a tensor structure.

As in the approach of~\cite{CcapaTtira:2011ga}, we define the system in terms of its Euclidean
action, ${\mathcal S}$.  Denoting by $A$ the Abelian gauge field, that
action may be written as follows:
\begin{equation}\label{eq:defaction}
	{\mathcal S}(A)\;=\;{\mathcal S}_0(A)\,+\,{\mathcal S}_{int}(A) \;,
\end{equation} 
where ${\mathcal S}_0(A)$ denotes the free gauge field action and
${\mathcal S}_{int}(A)$ the term that accounts for the coupling to the
mirrors.  
The former has the standard form:
\begin{equation}
{\mathcal S}_0(A)\,=\,\int d^4x \big( {\mathcal L}_{inv} + {\mathcal L}_{gf} \big)  
\end{equation}
with the gauge invariant piece: 
\begin{equation}
{\mathcal L}_{inv} \,=\, \frac{1}{4} F_{\mu\nu} F_{\mu\nu} \;,
\end{equation}
and for the gauge-fixing term we assume the form: 
\mbox{${\mathcal L}_{gf}=\frac{1}{2a}(\partial \cdot A)^2$}, with $a$ being a
positive real constant. 

The interaction action ${\mathcal S}_{int}$ is assumed to be composed of two
terms, each one describing the interaction between $A$ and a mirror:
\begin{equation}
	{\mathcal S}_{int} \;=\; {\mathcal S}_L \,+\, {\mathcal S}_R \;. 
\end{equation}
${\mathcal S}_I$ ($I = L,R)$, will be assumed to describe the interaction
with a single mirror, whose properties are time independent as well as
homogeneous and isotropic on each \mbox{$x_3={\rm constant}$} plane.
Regarding the $x_3$ direction (normal to both mirrors), we assume the
properties of the mirrors to be local functions of that coordinate. 

Besides, we use the fact that the interaction terms preserve gauge
invariance. This is guaranteed, if the current due to the charged
microscopic degrees of freedom which induce the coupling terms is
conserved.  Finally, the coupling terms are assumed to be quadratic in
$A_\mu$, which is a reasonable assumption to make when one deals with media
that may be appropriately described by linear response theory.

Then $S_I$ may be put into a more explicit form: using a
shorthand notation for the integrations, and assuming the $I$ mirror to be
centered at $x_3 = a_I$, we may write the term that describes its interaction
with the gauge field as follows: 
\begin{equation}\label{eq:defsiagen}
	{\mathcal S}_I(A) \;=\; \frac{1}{2} \int_{x_\parallel, x'_\parallel,
	x_3} A_{\mu}(x_\parallel,x_3) 
\Pi^{(I)}_{\mu \nu}(x_\parallel,x'_\parallel;x_3 -a_I) 
A_{\nu}(x'_\parallel,x_3) \;,
\end{equation}
where $\Pi^{(I)}_{\mu \nu} = \langle J_\mu J_\nu \rangle$ is the vacuum
polarization tensor, i.e., the correlator between currents, for the
matter fields on the $I$ mirror. 

Equation (\ref{eq:defsiagen}) suggests the consideration of two 
situations, the second a particular case of the first, regarding the
mirror's extent along the normal coordinate.
Firstly, we may regard it to have small width, in the sense that the
charge carriers in the medium are strongly concentrated in a finite $x_3$
region.  Since there is no current along $x_3$, the vacuum polarization
tensor (a correlator between currents) will be zero when one or two 
of its indices equals $3$. 
Secondly, we shall deal with the zero-width limit of the previous case. 

Here, the currents are essentially planar, and we shall then
neglect the action of $\Pi^{(I)}_{\mu \nu}$ on the third component
of the gauge field. 

Thus, in the small width case we shall have, 
\begin{equation}\label{eq:defsia}
	{\mathcal S}_I(A) \;=\; \frac{1}{2} \int_{x_\parallel, x'_\parallel,
	x_3} A_{\alpha}(x_\parallel,x_3) 
\Pi^{(I)}_{\alpha \beta}(x_\parallel-x'_\parallel;x_3 -a_I) 
A_{\beta}(x'_\parallel,x_3)
\end{equation}
where $\Pi_{\alpha \beta}^{(I)}$ is the vacuum polarization tensor for the
medium confined to the $I$ mirror.  
A convention we use is that in (\ref{eq:defsia}), $\alpha$ and $\beta$ run from $0$ to $2$. This implies
that the mirrors shall only involve the parallel components of the electric
field, $E_{\parallel}$ and the normal component of the magnetic field,
$B_3$. 

The tensor $\Pi^{(I)}_{\alpha\beta}(y_0,{\mathbf y}_\parallel;x_3)$, ($y\equiv
x-x'$) is assumed to be, as a function of $x_3$,
concentrated on a region centered around $x_3=0$. Note that we are not
assuming that $\Pi^{(I)}_{\alpha\beta}(y_{\parallel};x_3)$ necessarily can be
written as the product of a function of $x_3$ by a function of
$y_\alpha$, $\alpha=0,1,2$. For the case of very thin slabs, like the ones we shall consider
when dealing with graphene-like mirrors, that factorization is a natural
assumption to make. However, one could consider vacuum polarization tensors which properties
depends non trivially on the normal coordinate.

Performing a partial Fourier transformation in (\ref{eq:defsia}), i.e.,
just for the time and the parallel coordinates, we see that: 
\begin{equation}
{\mathcal S}_I(A) \;=\; \frac{1}{2} \int_{k_{\parallel},x_3}
\tilde{A}^*_{\alpha}(k_{\parallel},x_3) \,
\widetilde{\Pi}^{(I)}_{\alpha \beta}(k_{\parallel},x_3-a_I) \,
\tilde{A}_{\beta}(k_\parallel,x_3) \;.
\end{equation}
Here, and in what follows, we use the notation $k_\parallel \equiv
(k_0,k_1,k_2) = (k_0,{\mathbf k}_\parallel)$.  We
implicitly assume that the $k_0$ component is summed over discrete values,
$k_0 = \omega_n = \frac{2 \pi n}{\beta}$ (the Matsubara frequencies) at
finite temperature, and integrated (continuum values) at zero temperature. 

We have thus set up the general structure of the kind of systems that we
shall consider here. In the next section we show how to decompose the
problem of evaluating $\Gamma_C$ for the gauge field into two independent
one-dimensional systems, each one corresponding to a single real scalar field.
%%%%%%%%%%%%%%%%%%%%%%%%%%%%%%%%%%%%%%%%%%%%%%%%%%%%%%%%%%%%%%%%%%%%%%%%%%%%%%%
%%%%%%%%%%%%%%%%%%%%%%%%%%%%%%%%%%%%%%%%%%%%%%%%%%%%%%%%%%%%%%%%%%%%%%%%%%%%%%%
%%%%%%%%%%%%%%%%%%%%%%%%%%%%%%% Reduction %%%%%%%%%%%%%%%%%%%%%%%%%%%%%%%%%%%%%
%%%%%%%%%%%%%%%%%%%%%%%%%%%%%%%%%%%%%%%%%%%%%%%%%%%%%%%%%%%%%%%%%%%%%%%%%%%%%%%
%%%%%%%%%%%%%%%%%%%%%%%%%%%%%%%%%%%%%%%%%%%%%%%%%%%%%%%%%%%%%%%%%%%%%%%%%%%%%%%
\section{Reduction to one-dimensional systems}\label{sec:reduction}
Thus each mirror has been characterized by its vacuum polarization tensor
$\widetilde{\Pi}^{(I)}$. It is convenient to decompose each one of them in
terms of scalar functions, something that can be achieved, for example, by
expanding the tensor into a complete set of orthogonal projectors. That
decomposition is rather general, since it can be obtained as a consequence
of the assumptions we have made. 

Let us first note that, current conservation of the charge carriers in the
media implies that, for each $x_3$, the tensor
$\widetilde{\Pi}^{(I)}_{\alpha \beta}$ is transverse, namely: 
\begin{equation}\label{eq:ward}
k_{\alpha}\widetilde{\Pi}_{\alpha \beta}^{(I)}\;=\;0\;.
\end{equation}
Regarding the condition above, we can find two independent solutions to the
transversality condition, so that $\widetilde{\Pi}^{(I)}_{\alpha \beta}$
may be decomposed into two irreducible transverse tensors (projectors), in
terms of two scalars. Indeed, the assumed isotropy and homogeneity of the
media along the parallel directions, means that we can construct two
independent transverse tensors using as building blocks the elements:
$\breve{k}_{\alpha}\equiv k_{\alpha} - k_0 n_{\alpha}$, and
$\breve{\delta}_{\alpha \beta} \equiv \delta_{\alpha \beta} -
n_{\alpha}n_{\beta}$, where $n=(1,0,0)$. Note that the presence of $n$ is
allowed since Poincar\'e invariance on the $x_3=0$ spacetime does not hold
necessarily true.   

Two independent projectors ${\mathcal P}^{t}$ and ${\mathcal P}^{l}$ that are
solutions of (\ref{eq:ward}) may be written as follows:
\begin{equation}
	{\mathcal P}^{t}_{\alpha \beta} \equiv {\breve{\delta}_{\alpha \beta}} - \frac
{\breve{k}_{\alpha} \breve{k}_{\beta}}{{\breve{k}}^2}
\end{equation}
and
\begin{equation}
	{\mathcal P}^{l}_{\alpha \beta} \equiv {\mathcal P}^{\perp}_{\alpha
	\beta} - {\mathcal P}^{t}_{\alpha \beta} 
\end{equation}
where
\begin{equation}
P^{\perp}_{\alpha \beta} \equiv \delta_{\alpha \beta} - \frac
{k_{\alpha}k_{\beta}}{k_\parallel^2} \;
\end{equation}
is the transverse projector corresponding to a $2+1$ dimensional Poincar\'e
covariant theory.  For the sake of completeness, we also introduce the
`parallel' projector
${\mathcal Q}$:
\begin{equation}
{\mathcal Q}_{\alpha\beta} \,\equiv\, \frac{k_\alpha
k_\beta}{k_\parallel^2} \;.
\end{equation}
They satisfy the following algebraic properties:
	$$
	{\mathcal P}^\perp + {\mathcal Q} = I \;,\;\;
	{\mathcal P}^t + {\mathcal P}^l = {\mathcal P}^\perp 
	$$
	$$
	{\mathcal P}^t {\mathcal P}^l = {\mathcal P}^l {\mathcal P}^t = 0
	\;,\;\; {\mathcal Q} {\mathcal P}^t = {\mathcal P}^t {\mathcal Q} = 0 
	\;,\;\; {\mathcal Q} {\mathcal P}^l = {\mathcal P}^l {\mathcal Q} = 0 
	$$
\begin{equation}
\big({\mathcal P}^\perp\big)^2 = {\mathcal P}^\perp \;,
\big({\mathcal Q})^2 = {\mathcal Q} \;,
\big({\mathcal P}^t\big)^2 = {\mathcal P}^t \;,
\big({\mathcal P}^l\big)^2 = {\mathcal P}^l \;,
\end{equation}
where $I_{\alpha\beta} = \delta_{\alpha\beta}$.
Therefore we can express $\widetilde{\Pi}_{\alpha \beta}^{(I)}$ as follows:
\begin{equation}\label{polarizationdelvacio}
\tilde{\Pi}_{\alpha \beta}^{(I)}(k_\parallel, x_3) \;=\; f^{(I)}_t ({k}^{2}_{0},
\mathbf{k}_\parallel^{2}, x_3) \, 
{\mathcal P}^{t}_{\alpha \beta} + f^{(I)}_l(k^{2}_{0},\mathbf{k}_\parallel^{2}, x_3)
{\mathcal P}^{l}_{\alpha \beta} \;.	
\end{equation}
In this way, we have succeeded in characterizing the $I$ mirror by two
functions, $f^{(I)}_{t,l}$. To proceed to the reduction of the problem of
evaluating ${\mathcal Z}(\beta)$ to  one-dimensional functional
determinants, we shall perform the same Fourier transformation we used for the
interaction terms, for the free action ${\mathcal S}_0$. Adopting the
Feynman ($a \equiv 1$) gauge choice,
\begin{equation} 
{\mathcal S}_0 = \frac{1}{2} \int d^4x \, A_\mu(x) (- \partial^2) A_\mu(x) 
\end{equation}
we see that
\begin{eqnarray} 
{\mathcal S}_0 &=& \frac{1}{2} \int_{k_{\parallel},x_3} \big[
	\widetilde{A}^*_\alpha(k_\parallel,x_3) (-\partial_3^2 +
k_\parallel^2) \widetilde{A}_\alpha(k_\parallel,x_3) \nonumber\\
&+&\widetilde{A}^*_3(k_\parallel,x_3) (-\partial_3^2 +
k_\parallel^2) \widetilde{A}_3(k_\parallel,x_3) \Big]\,.
\end{eqnarray}

Then, the complete action ${\mathcal S}$ may be split into two terms, one
depending on $\tilde{A}_\parallel \equiv (\tilde{A}_\alpha)$ 
and the other on $\tilde{A}_3$:
\begin{equation}\label{eq:actdec} 
{\mathcal S} \;=\; {\mathcal S}_\parallel(\tilde{A}_\parallel) \,+\, 
{\mathcal S}_3(\tilde{A}_3) \;,
\end{equation}
with:
\begin{eqnarray}
 {\mathcal S}_\parallel = \frac{1}{2} \int_{k_{\parallel},x_3} 
\widetilde{A}^*_\alpha(k_\parallel,x_3) \big[ (-\partial_3^2 +
k_\parallel^2 ) \delta_{\alpha\beta} \nonumber\\
+ \sum_{I} \widetilde{\Pi}_{\alpha \beta}^{(I)}(k_\parallel, x_3 - a_I)\big] 
\widetilde{A}_\beta(k_\parallel,x_3) 
\end{eqnarray}
and
\begin{equation}
 {\mathcal S}_3 \,=\, \frac{1}{2} \int_{k_{\parallel},x_3} 
\widetilde{A}^*_3(k_\parallel,x_3) (-\partial_3^2 +
k_\parallel^2 ) \widetilde{A}_3(k_\parallel,x_3) \;.
\end{equation}

Note that, because of (\ref{eq:actdec}), and the fact that ${\mathcal S}_3$
does not involve any coupling to the mirrors, we may write the ratio 
between ${\mathcal Z}(\beta)$ and ${\mathcal Z}_0(\beta)$ as follows:
\begin{equation}
\frac{{\mathcal Z}(\beta)}{{\mathcal Z}_0(\beta)} \,=\, 
\frac{{\mathcal Z}_\parallel(\beta)}{{\mathcal Z}_{\parallel 0}(\beta)}  
\end{equation}
with:
\begin{equation}\label{eq:defzetapar}
{\mathcal Z}_\parallel(\beta) \;=\; \int {\mathcal D}\tilde{A}_\parallel \,
e^{-{\mathcal S}_\parallel({\tilde A}_\parallel)} \;.
\end{equation}

Applying the properties satisfied by the projectors, we see that:
\begin{equation}
\delta_{\alpha\beta} \;=\; {\mathcal P}^t_{\alpha\beta} \,+\, {\mathcal P}^l_{\alpha\beta} \,+\,
{\mathcal Q}_{\alpha\beta} 
\end{equation}
which allows us to write:
\begin{eqnarray}
	{\mathcal S}_\parallel &=& \frac{1}{2} \int_{k_{\parallel},x_3} 
\widetilde{A}^*_\alpha(k_\parallel,x_3) \Big\{ \big[-\partial_3^2 +
k_\parallel^2 +  \sum_I f^{(I)}_t ({k}^{2}_{0}, \mathbf{k}_\parallel^{2}, x_3 - a_I) 
\big] {\mathcal P}^{t}_{\alpha \beta} \nonumber\\
&+&  \big[-\partial_3^2 + k_\parallel^2 +  \sum_I f^{(I)}_l ({k}^{2}_{0},
\mathbf{k}_\parallel^{2}, x_3 -a_I) \big] {\mathcal P}^{l}_{\alpha \beta} \Big\} 
\widetilde{A}_\beta(k_\parallel,x_3) \;,
\end{eqnarray}
what concludes the reduction. Indeed, note that the action has been reduced
to a quadratic form  for an operator which has been decomposed into orthogonal
rank-one projectors.

\section{Lifshitz formula}\label{sec:Lifshitz}
To obtain the Lifshitz formula for this kind of model, we proceed
as follows: In the path integral for ${\mathcal Z}_\parallel$, we may
decompose the gauge field:
\begin{equation}
\tilde{A}_\parallel = \tilde{A}^{(t)} + {\tilde A}^{(l)} 
\end{equation}
$\tilde{A}^{(t,l)} \equiv {\mathcal P}^{(t,l)} \tilde{A}_\parallel$ 
under which the path integral measure factorizes. Thus,
\begin{equation}
{\mathcal Z}_\parallel(\beta) \;=\;  {\mathcal Z}^{(t)}(\beta) \, {\mathcal Z}^{(l)}(\beta) 
\end{equation}
where each factor is obtained as the result of performing a functional
integral over one scalar degree of freedom, namely,
\begin{equation}\label{eq:defzet}
	{\mathcal Z}^{(t,l)}(\beta) \,=\, \int {\mathcal D}\tilde{A}^{(t,l)} \,
	\exp\big\{-{\mathcal S}^{(t,l)}({\tilde A}^{(t,l)})\big\} \;
\end{equation}
where 
\begin{eqnarray}
{\mathcal S}^{(t,l)}({\tilde A}^{(t,l)})&=&\frac{1}{2} \int_{k_{\parallel},x_3} 
\widetilde{A}^{*(t,l)}_\alpha(k_\parallel,x_3) \big[-\partial_3^2 +
k_\parallel^2  \nonumber \\
&+& \sum_I f^{(I)}_{t,l} ({k}^{2}_{0}, \mathbf{k}_\parallel^{2}, x_3 - a_I) 
\big]
\widetilde{A}^{(t,l)}_\alpha(k_\parallel,x_3) \;.
\end{eqnarray}
Then we see that the free energy becomes:
\begin{equation}
	\Gamma_C(\beta) \,=\,\Gamma_t(\beta) + \Gamma_l(\beta)  
\end{equation}
where
\begin{equation}
	\Gamma_{t,l}(\beta) \,= - \, \frac{1}{\beta} \, \lim_{L\to \infty} 
	\Big[ \frac{1}{L^2} \, \log \frac{{\mathcal
	Z}^{(t,l)}(\beta)}{{\mathcal Z}^{(t,l)}_0(\beta)}\Big]
\end{equation}
or
\begin{equation}
	\Gamma_{t,l}(\beta) \,= \, \frac{1}{2} \,\int
	\frac{d^3k_{\parallel}}{(2 \pi)^3}\,
	\log \Big[\frac{\det \widetilde{T}_{t,l}(k_\parallel)}{\det \widetilde{T}_0(k_\parallel)}\Big] \;
\end{equation}
where:
\begin{eqnarray}
	\widetilde{T}_{t,l}(k_\parallel) &=& -\partial_3^2 + k_\parallel^2 +
	\widetilde{V}_{t,l}(x_3,k_\parallel)  \nonumber\\
\widetilde{T}_0(k_\parallel) &=& -\partial_3^2 + k_\parallel^2 \;,
\end{eqnarray}	
and:	
\begin{equation}	
	\widetilde{V}_{t,l}(x_3,k_\parallel) \,=\, \sum_I f^{(I)}_{t,l} ({k}^{2}_{0}, 
	\mathbf{k}_\parallel^{2}, x_3 - a_I) \;.
\end{equation}
The system has been reduced to two independent Casimir problems, each one
of them corresponding to a real scalar field in the presence of its
potential background $\widetilde{V}_{t,l}$.  These potentials are built in
terms of the functions that appear in the decomposition of the vacuum
polarization tensor into a set of irreducible tensors.

Applying the general formula derived in~\cite{CcapaTtira:2011ga}, we may write for
each contribution above:
\begin{equation}\label{eq:resgtl}
	\Gamma_{t,l}(\beta) \,= \, \frac{1}{2} \, \int_{k_\parallel} 
	\log \left[ 1 + \frac{T^{(2)}_{12}}{T^{(1)}_{11}}
		\frac{T^{(2)}_{21}}{T^{(1)}_{11}} e^{- 2\,
	\mid k_\parallel\mid \, l} \right]_{t,l}\;,
\end{equation}
where $T_{t,l}$ is the result of performing the following change of basis to the matrix $A_{t,l}$:
\begin{equation}
T_{t,l}= B^{-1}A_{t,l}B
\end{equation}
with
\begin{equation}
B=\frac{1}{\sqrt2}
\left( \begin{array}{rr}
1 & 1 \\
1 & -1
\end{array} \right),
\end{equation}
and $A_{t,l}$ are defined as in~\cite{CcapaTtira:2011ga},
regarding each one, $t$ or $l$, as due to an independent field, in its own
background potential.

\section{Zero width mirrors}\label{sec:zerowidth}
We characterize thin mirrors here as systems where the interaction between
field and mirrors is confined to zero-width planes. Thus, in this case,
\begin{equation}
	f^{(I)}_{t,l}({k}^{2}_{0}, \mathbf{k}^{2}, x_3 -
	a_I) \;=\; \delta(x_3 - a_I) \; g^{(I)}_{t,l}({k}^{2}_{0},
	\mathbf{k}^{2})  \,,
\end{equation}
and
\begin{equation}	
	\widetilde{V}_{t,l}(x_3,k_\parallel) \;=\; 
	\sum_I \delta(x_3 - a_I) \; g^{(I)}_{t,l}({k}^{2}_{0},
	\mathbf{k}_\parallel^{2}) \;.
\end{equation}

Recalling the known result of~\cite{CcapaTtira:2011ga} for the case of a
real scalar field in the presence of zero width mirrors, we see that:
\begin{equation}
\Gamma_{t,l}(\beta)=
\frac{1}{2} 
\int_{k_\parallel} 
\log \left\{ 1 - 
	\frac{g^{(L)}_{t,l}({k}^{2}_{0}, \mathbf{k}_\parallel^{2}) 
	g^{(R)}_{t,l}({k}^{2}_{0}, \mathbf{k}_\parallel^{2}) e^{- 2 |k_\parallel| l}}{\big[ 2 |k_\parallel| +
	g^{(L)}_{t,l}({k}^{2}_{0}, \mathbf{k}_\parallel^{2})\big] 
	\big[ 2 |k_\parallel| + g^{(R)}_{t,l}({k}^{2}_{0}, \mathbf{k}_\parallel^{2})\big]}\right\}.
\end{equation}

Then, the Casimir force per unit area becomes:
\begin{equation}
{\mathcal F}_C(\beta)={\mathcal F}_C^{(t)}(\beta)+{\mathcal F}_C^{(l)}(\beta)
\end{equation}
with
\begin{equation}
{\mathcal F}_C^{(t,l)}(\beta)= - 
\int_{k_\parallel}
	\frac{|k_\parallel|g^{(L)}_{t,l}
	g^{(R)}_{t,l} e^{- 2 |k_\parallel| l}}
{( 2 |k_\parallel| +
	g^{(L)}_{t,l}) 
	( 2 |k_\parallel| + g^{(R)}_{t,l})-g^{(L)}_{t,l} 
	g^{(R)}_{t,l} e^{- 2 |k_\parallel| l}}\;,
\end{equation}
where the arguments of $g^{(L)}_{t,l}({k}^{2}_{0}, \mathbf{k}_\parallel^{2})$ and 
$g^{(R)}_{t,l}({k}^{2}_{0}, \mathbf{k}_\parallel^{2})$ were omitted.

For a graphene sheet
(\cite{Bordag:2006zc},\cite{Bordag:2009fz},\cite{Fialkovsky:2011pu}), which
can be reasonably described by a zero-width mirror, the corresponding $g$
functions may be read off from its vacuum polarization tensor, the result being:
\begin{eqnarray}
g_t(k^2_0, \mathbf{k}_\parallel^{2}) &=& \alpha	\sqrt{k_0^2 + v_F^2 {\mathbf k}^2} \nonumber\\
g_l({k}^{2}_{0}, \mathbf{k}_\parallel^{2}) &=& \alpha \frac{k_0^2 +{\mathbf k}^2}{\sqrt{k_0^2 + v_F^2 {\mathbf k}^2}}
\end{eqnarray}
with $\alpha=\frac{e^2 N}{16}$, where $N$ is the number of fermion flavours, $e$ the couppling
constant, and $v_F$ the Fermi velocity.

Using these expressions into the general formula for thin mirrors, we
obtain the Casimir force for cases involving either two graphene sheets or,
as a limiting case, a graphene sheet and a conducting mirror. The latter
may be obtained from the graphene case by setting the Fermi velocity to $1$
and $\alpha \to \infty$ in one of the mirrors. 

\begin{center}
\begin{figure}
\begin{picture}(0,0)
\includegraphics{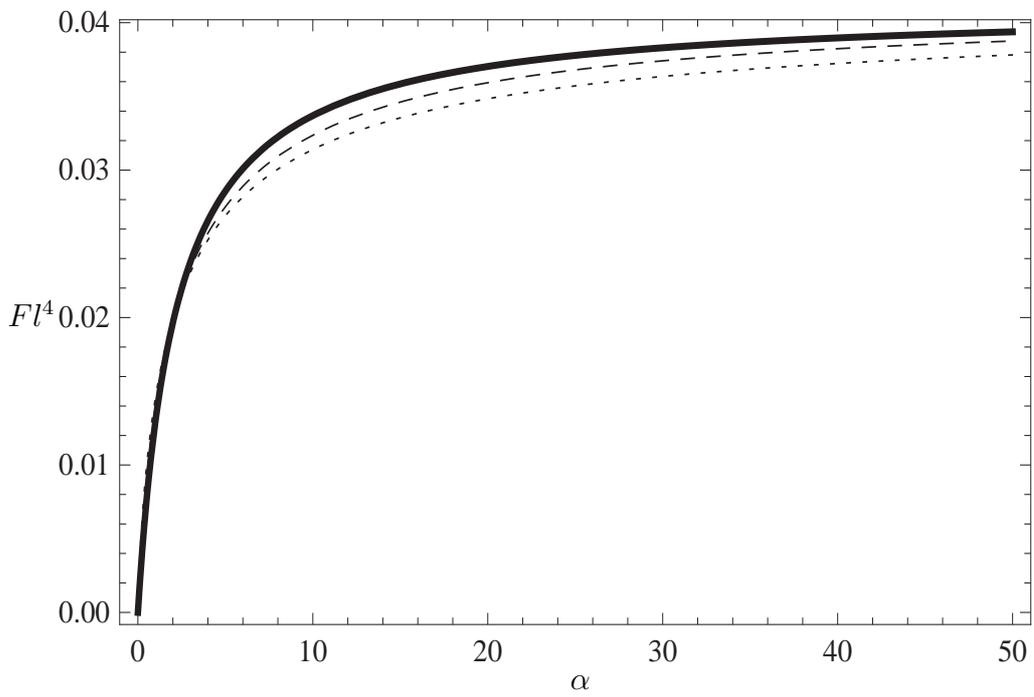}%
\end{picture}%
\setlength{\unitlength}{4144sp}%
\begingroup\makeatletter\ifx\SetFigFont\undefined%
\gdef\SetFigFont#1#2#3#4#5{%
  \reset@font\fontsize{#1}{#2pt}%
  \fontfamily{#3}\fontseries{#4}\fontshape{#5}%
  \selectfont}%
\fi\endgroup%
\begin{picture}(6945,4263)(256,-4324)
\put(0,-2230){\makebox(0,0)[lb]{\smash{{\SetFigFont{12}{14.4}{\rmdefault}{\mddefault}{\updefault}{\color[rgb]{0,0,0}$F l^4$}%
}}}} 
\put(3330,-4400){\makebox(0,0)[lb]{\smash{{\SetFigFont{12}{14.4}{\rmdefault}{\mddefault}{\updefault}{\color[rgb]{0,0,0}$\alpha$}%
}}}}
\end{picture}%
\caption{Casimir force times $l^4$ as a function of $\alpha$, for a
perfectly conducting and a graphene mirror with different values of $v_F$. 
The solid line corresponds to $v_F=1$, the dashed line to $v_F=0.2$ and the dotted one to $v_F=0$.}\label{fig:fig1}
\end{figure}
\end{center}

\begin{center}
\begin{figure}
\begin{picture}(0,0)%
\includegraphics{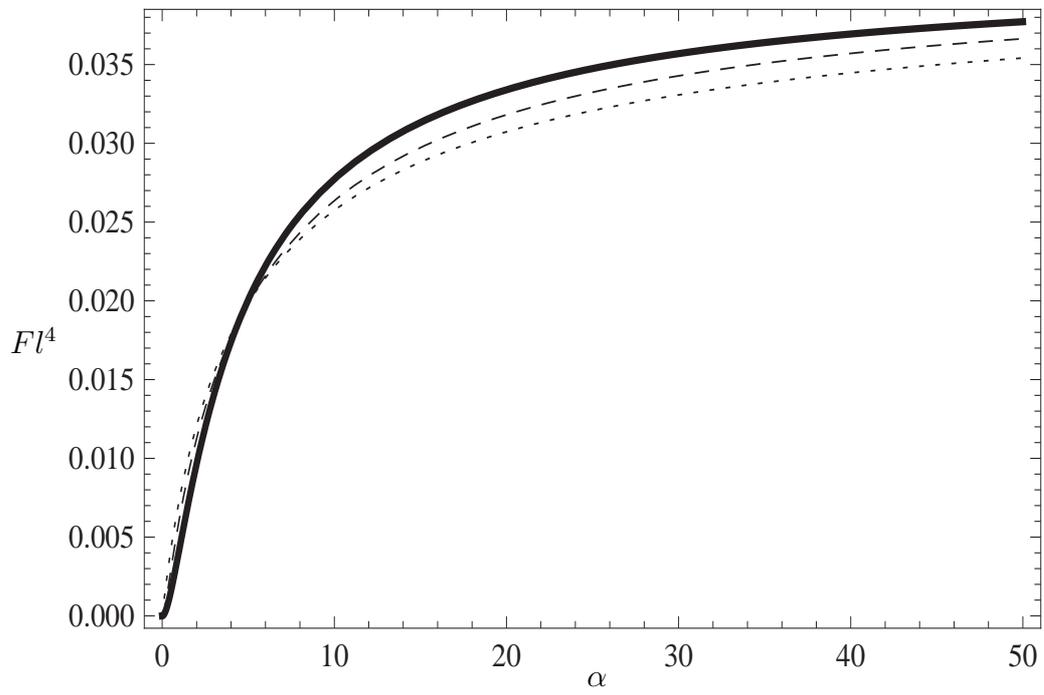}%
\end{picture}%
\setlength{\unitlength}{4144sp}%
\begingroup\makeatletter\ifx\SetFigFont\undefined%
\gdef\SetFigFont#1#2#3#4#5{%
  \reset@font\fontsize{#1}{#2pt}%
  \fontfamily{#3}\fontseries{#4}\fontshape{#5}%
  \selectfont}%
\fi\endgroup%
\begin{picture}(6900,4218)(301,-4279)
\put(0,-2310){\makebox(0,0)[lb]{\smash{{\SetFigFont{12}{14.4}{\rmdefault}{\mddefault}{\updefault}{\color[rgb]{0,0,0}$F l^4$}%
}}}}
\put(3415,-4300){\makebox(0,0)[lb]{\smash{{\SetFigFont{12}{14.4}{\rmdefault}{\mddefault}{\updefault}{\color[rgb]{0,0,0}$\alpha$}%
}}}}
\end{picture}%
\caption{Casimir force times $l^4$ as a function of $\alpha$, for two identical 
graphene mirrors characterized by $v_F$. The solid line corresponds to $v_F=1$, 
the dashed line to $v_F=0.2$ and the dotted one to $v_F=0$.}\label{fig:fig2}
\end{figure}
\end{center}

In figure~\ref{fig:fig1} we plot the zero temperature pressure times $l^4$
as a function of $\alpha$ for the case of a perfectly conducting mirror in
front of a graphene sheet, for different values of $v_F$, and in
figure~\ref{fig:fig2} for two identical graphene sheets. 
Note that in both figures the solid line corresponding to $v_F=1$ represents 
a `relativistic matter' case,  where $f^{(I)}_{t}=f^{(I)}_{l}$, considered
in~\cite{Fosco:2008td}.

\newpage
\section{Conclusions}\label{sec:conclusions}
We have derived a general expression for the Casimir free energy, using an
entirely field theoretic approach, whereby the problem is analyzed in terms
of the functional determinant for a fluctuating Abelian gauge field.
We have shown that, under some assumptions regarding form of the coupling
between the gauge field and the mirrors, the problem can be reduced to
scalar systems, for which one can apply the previuosly known expression for
the functional determinant. 

The result is expressed in terms of the invariants of the Euclidean version
of the vacuum polarization tensor due to the charged matter inside the
mirror. In this way one may bypass the calculation of the reflection
coefficients of each mirror, as it would be the case with the usual version
of Lifshitz formula. Besides, the result for small-width mirrors allows for
 cases where the material media have a non trivial dependence along the
 normal direction; for example, one could consider vacuum polarization
 tensors corresponding to stratified media.

For zero width mirrors with graphene like properties, we have shown that
the QFT approach yields results which are  consistent with the ones
presented in \cite{Bordag:2006zc},\cite{Bordag:2009fz},\cite{Fialkovsky:2011pu}.

\section*{Acknowledgements}
This work was supported by CONICET, and UNCuyo.

\end{document}